\documentclass{elsart}
\usepackage{graphics}

\usepackage{epsfig}

\usepackage{amssymb}

\usepackage[english,francais]{babel}

\def\np#1#2#3{Nucl. Phys. {\bf B#1} (#2) #3}
\def\pl#1#2#3{Phys. Lett. {\bf #1B} (#2) #3}

\def\cmp#1#2#3{Comm. Math. Phys. {\bf #1} (#2) #3}

\def\p{\partial}
\def\pb{\bar{\partial}}
\def\dir{{\CD}\hskip -7pt \slash \hskip 6pt}
\def\dd{{\rm d}}
\def\ib{\bar{i}}

\def\Tr{{\rm Tr}}

\def\Det{{\rm Det}}

\def\a{{\alpha}}

\def\p{{\partial}}
\def\b{{\beta}}
\def\d{{\delta}}

\def\g{{\gamma}}
\def\e{{\epsilon}}

\def\ve{{\varepsilon}}
\def\vf{{\varphi}}
\def\m{{\mu}}

\def\u{{\Upsilon}}
\def\l{{\lambda}}
\def\s{{\sigma}}

\def\o{{\omega}}

\chardef\tempcat=\the\catcode`\@ \catcode`\@=11
\def\cyracc{\def\u##1{\if \i##1\accent"24 i%
    \else \accent"24 ##1\fi }}
\newfam\cyrfam
\font\tencyr=wncyr10
\def\cyr{\fam\cyrfam\tencyr\cyracc}

\def\CA{{\mathcal A}}

\def\CD{{\mathcal D}}

\def\CF{{\mathcal F}}

\def\CH{{\mathcal H}}
\def\CI{{\mathcal I}}

\def\CL{{\mathcal L}}
\def\CM{{\mathcal M}}
\def\CN{{\mathcal N}}
\def\CO{{\mathcal O}}
\def\CP{{\mathcal P}}

\def\CS{{\mathcal S}}

\def\CV{{\mathcal V}}

\def\CZ{{\mathcal Z}}

\def\bA{{\bf A}}
\def\bB{{\bf B}}
\def\bC{{\bf C}}

\def\bH{{\bf H}}

\def\bM{{\bf M}}

\def\bP{{\bf P}}
\def\bR{{\bf R}}
\def\bS{{\bf S}}
\def\bT{{\bf T}}

\def\bZ{{\bf Z}}

\newtheorem{e-proposition}[theorem]{Proposition}

\newtheorem{e-definition}[theorem]{Definition\rm}

\begin{document}

\begin{frontmatter}
\selectlanguage{english}
\title{{\bZ}-theory: chasing ${\mathfrak m/f}$ theory}

\vspace{-3.6cm}

\selectlanguage{francais}
\title{A la recherche de la ${\mathfrak m}$-th\'eorie  perdue}

\vspace{1cm}
\selectlanguage{english}
\author[NN]{Nikita Nekrasov}
\ead{nikita@ihes.fr}
\address[NN]{Institut des Hautes Etudes Scientifiques, 91440 Bures-sur-Yvette France}
\begin{abstract}
We present the evidence for the existence of the topological string analogue
of M-theory, which we call {\bZ}-theory. The corners of {\bZ}-theory moduli space
correspond to the Donaldson-Thomas theory, Kodaira-Spencer theory,
Gromov-Witten theory, and Donaldson-Witten theory.
We discuss the relations
of
{\bZ}-theory with Hitchin's gravities in six and seven
dimensions, and make our own proposal, involving spinor generalization of Chern-Simons theory
of three-forms.
{\it To cite this article:
N.~Nekrasov
, C. R.
Physique 4 (2004).}

\vskip 0.5\baselineskip

\selectlanguage{francais}
\noindent{\bf R\'esum\'e}
\vskip 0.5\baselineskip
\noindent
 {\it Pour citer cet article~: N.~Nekrasov, C. R.
Physique 4 (2004).}
\vskip 0.5\baselineskip
\noindent{\small{\it Mots-cl\'es~:} corde topologique; M theorie; gravitation
quantique}}
\end{abstract}
\end{frontmatter}

\selectlanguage{english}
\section{Introduction}

The past ten years of string theory development have taught us that string
theory is a wrong name for the fundamental theory of quantum gravity.
We know that the theory has a moduli space of vacua, that this moduli space
has some singularities, and we know that the expansion near different
singularities look like different string theories, or like eleven
dimensional supergravity \cite{wittenm}.
In the context of topological strings the situation used to be different,
but recent advances in this field suggest the similar picture.
In the past few months a few striking conjectures have been put forward
concerning the strong-weak dualities, relating topological strings of {\bA} and
{\bB} types on the same Calabi-Yau three-fold $X$. The conjectures relate the
perturbative type {\bA} string calculations to the D-brane {\bB}
type calculations,
and vice versa. So far most of the known checks of this S-duality conjecture
involved only {\bB}-type branes.

Strong-weak coupling duality in the physical superstring follows from
the existence of some
higher-dimensional theory, such that its compactification on tori
gives rise to the dual theories.
The purpose of this lecture is to draw a similar picture of what we called
at various occasions (M)athematical M-theory, $\mathfrak m$,  or $\mathfrak f$-theory.
Some
people call it topological M-theory \cite{dijkgraafstring}, \cite{grassivanhove}. Since for us the main object is a
certain partition function, which we denote by $Z$, in this paper
we shall call the
missing theory the $\bZ$-theory.

The simplest idea would be that
the physical $M$-theory, whatever it is, is related to $\bZ$-theory, just
like physical strings are related to the topological strings \cite{bcov}, \cite{agnt}. This may well
be true, but two warning signs are in order: this relation would not explain
the relation between the topological gauge theory on ${\bR}^4$ and the
topological string on local Calabi-Yau manifold within {\bZ}-theory; to
actually engineer the relation between the ${\CN}=1$ theories in four
dimensions (which is what one gets by compactifying {\bM}-theory on
${\CZ}_7$) and the topological strings one has to use CY compactifications
with fluxes (which could be in principle related to $G_2$-compactifications, but this makes the whole
construction less pretty) \cite{dv}.


\section{Evidence for {\bZ}-theory}
\label{}

In this section we describe briefly the topological string theory and the
topological gauge theory computations which correspond to various
degenerations of {\bZ}-theory.

\subsection{{\bA} story}

\noindent
{\bf Gromov-Witten corner.}
Consider closed {\bA} type topological string on Calabi-Yau threefold $X$.
Let $k$ denote the (complexified) Kahler form of $X$, and ${\mathfrak t} = [ k ]
\in H^{2}(X, {\bC})$.
The partition function is defined as a formal series in the string coupling
constant $\hbar$:
\begin{equation}
Z_{A}(X, {\mathfrak t}; {\hbar})^{GW} = {\exp} \sum_{g=0}^{\infty}
{\hbar}^{2g-2} {\CF}_{g}(X; {\mathfrak t})
\label{topa}
\end{equation}
where
\begin{equation} \label{fga}
{\CF}_{g}(X; {\mathfrak t}) = \sum_{{\b} \in H_{2}(X; {\bZ})}
{\exp}\left({-\int_{\b} {\mathfrak t}}\right) N_{g}({\b})
\end{equation}
and $N_{g}({\b})$ is the ``number'' of genus $g$ stable holomorphic maps to
$X$ which land in the homology class $\b$. The word ``number'' here can be
defined more precisely using the virtual fundamental cycles but we shall not
do that.

\noindent{\bf Lagrangian branes.} By definition (\ref{topa}) the partition function is perturbative in $\hbar$. The
relation to physical superstring \cite{bcov} suggests that there are
non-perturbative corrections to the ``correct'' definition of $Z_{A}$.
These corrections, presumably, come from $D$-branes.
There are natural $D$-branes in the topological string context. Namely,
for any Lagrangian submanifold $L \subset X$ (where $X$ is viewed as
symplectic manifold), one can define the relative analogue of Gromov-Witten
theory, i.e. stable maps of Riemann surfaces with boundaries, which land on
$L$. Moreover, $L$ may have several components, each component may have
multiplicities and so on. In the most naive approach one would combine the
effects of closed strings and open strings as follows:
\begin{equation} \label{topaop}
Z_{A}^{?} (X; {\mathfrak t}, {\mathfrak s}; {\hbar} \vert {\Lambda} )
= Z_{A}(X; {\mathfrak t}; {\hbar})^{GW} \times \sum_{{\ell} \in {\Lambda} \subset
H_{3}(X, {\bZ})} {\exp}\left({- {1\over{\hbar}}\int_{{\ell}}{\mathfrak s}}\right)
{\CN}_{\ell}({\hbar})
\end{equation}
where ${\CN}_{\ell}({\hbar}) = \sum_{h \in {\bZ}} {\CN}_{{\ell},
h}{\hbar}^{2h-2}$
counts stable maps of (possibly disconnected) Riemann surfaces (hence the
total genus can be arbitrary), with boundaries, which land on the Lagrangian
submanifolds $L_i$, $i=1, \ldots , k_{\ell}$ which represent the homology
cycle $\ell$. The homology cycles must belong to a Lagrangian (with respect
to the intersection pairing) sublattice in $H_{3}(X, {\bZ})$\footnote{The
reason for taking only ``half'' of all possible 3-cycles is the
electro-magnetic duality of the effective four dimensional sugra}
If these Lagrangian submanifolds are not
simply-connected, then one modifies the definition of the numbers
${\CN}_{\ell, h}$ by considering the moduli spaces of the pairs $(L_i,
{\CL}_i)$ , where ${\CL}_{i}$ is the rank mult($L_i$) vector bundle on
$L_i$ with flat unitary connection. The stable map $({\Sigma}, {\p\Sigma})$
is weighted with the weight ${\Tr} P\exp\oint_{\p\Sigma} A$ where $A$ is the
pullback of the flat connection. The result it then somehow averaged over
the moduli space of the Lagrangian submanifolds with unitary flat bundles
over them. The logic of this construction is largely motivated by the
corresponding one on the {\bB} side.

We should learn from this discussion that although perturbative {\bA} string
only knows about the (complexified) symplectic structure of $X$, via $\mathfrak
t$-dependence, the non-perturbative corrections bring in extra structure,
the $3$-form $\mathfrak s$, which turns out (upon complexification again) to be
related to the $(3,0)$-form of the complex face of the Calabi-Yau manifold
$X$; there is a corresponding
term\footnote{
To arrive at the coupling (\ref{extrs}) we note that in the
presence of the boundary condition corresponding to $L$ the scalar fermions
${\psi}^{\m}$ in the worldsheet sigma model have zero modes corresponding to
the motion along $L$. The zero-observable ${\o}_{m_{1}\ldots
m_{p}}{\psi}^{m_{1}} \ldots {\psi}^{m_{p}}$
saturates these zero modes if $p = {\rm dim}L$
(so, in particular, in the more general setup for the type {\bA} topological
string one gets a similar coupling for the ${\half}{\rm dim}X$-forms).
The zero-observable inserted at the center of the disk breaks $SL_2$ down to
the compact subgroup $U(1)$. The one-point function is non-vanishing, since the
volume of $U(1)$ is finite.} in the target space theory action
\begin{equation} \label{extrs}
S_{\rm target} = \int_{L} {\mathfrak s} + k {\Tr}\left(
 A d A + {2\over 3} A^{3}\right)
\end{equation}

\subsection{{\bB} story}

Mirror symmetry relates type {\bA} string on $X$ to type {\bB} topological
string on $X^{\vee}$ - another Calabi-Yau manifold.
All the features of the type {\bA} string described above should be equally
present for {\bB} string
as well, order by order in $\hbar$. Indeed, mirror symmetry is the
equivalence of sigma models before their coupling to the two dimensional
gravity, and also holds in the presence of worldsheet boundaries.

\noindent {\bf Kodaira-Spencer corner.}
In particular, there exists a definition of the closed type $B$ partition
function, and \cite{bcov} suggests that it is given by some field
theory, i.e. instead of the integrals over the moduli spaces of Riemann
surfaces one works with the integrals over the moduli spaces of Riemann
graphs.  The classical, i.e. genus zero, definition of the type {\bB} string
free energy suggests an intimate connection to the symplectic geometry, via
the special geometry. The full partition function involves
$\hbar$-corrections, so the
story involves some sort of quantization of the symplectic manifold, which
gives rise to the special geometry, however
no satisfactory proposal about it has been put forward so far.
The {\it calibrated
CY manifold} is a pair: $(X^{\vee}, {\Omega})$, where $X^{\vee}$ is a complex
threefold, $K_{X^{\vee}} \approx {\CO}_{X^{\vee}}$ and $\Omega$ is nowhere
vanishing holomorphic $(3,0)$-form.
The moduli space ${\widetilde\CM}$ of calibrated CY manifolds has complex dimension
$1 + h^{2,1}_{X^{\vee}} = {\half}{\rm dim}H^{3}(X^{\vee}, {\bR})$. Moreover,
one can choose local coordinates on ${\widetilde\CM}$ to be the periods of
$\Omega$.
These periods are not independent: choose some basis of $A$ and
$B$ cycles in $H_{3}(X^{\vee}, {\bZ})$: $A^i \circ A^j = 0$, $A^i \circ B_j =
{\d}^{i}_{j}$, $B_{j} \circ B_i = 0$, $i = 0,1, \ldots, r$, $r = h^{2,1}(X^{\vee})$,
where $\circ$ stands for the
intersection index, and define:
\begin{equation} \label{perds}
t_i = \oint_{A^i} {\Omega}, \qquad t_{D}^{i} = \oint_{B_{i}}
{\Omega}
\end{equation}
Then $t^i$ are the local coordinates on $\widetilde\CM$ and
locally on $\widetilde\CM$ there exists a
holomorphic function ${\CF}_{0} = {\CF}_{0}(X^{\vee}; t)$ such that :
\begin{equation} \label{prep}
t_{D}^{i} = {{\p}{\CF}_0 \over {\p}t^i}
\end{equation}
This function, called prepotential, is the genus zero topological
{\bB} string partition function\footnote{More precisely, its third derivative
(in the special coordinates $t^i$) is the three-point function on a sphere
of the zero-observables ${\m}_{\ib}^{j}(x, {\bar x}) {\eta}^{\ib}{\theta}_{j}$ of {\bB} model, corresponding to the Beltrami
differentials}.
The topological {\bB} string couples naively only to the complex structure
deformations of $X^{\vee}$. However, it is well-known that the worldsheet
theory is anomalous, and the choice of $\Omega$ enters the definition of the
path integral measure.
The moduli space $\widetilde\CM$ is a cone over the moduli space $\CM$ of
complex structures of $X^{\vee}$. The rescaling of
$\Omega$
does not change the complex structure of $X^{\vee}$, so that the quotient
by this ${\bC}^*$ action gives $\CM$.
This ${\bC}^{*}$-action scales simultaneously $t_i$ and $t_{D}^i$, which
means that ${\CF}_0$ should be a homogeneous function of degree $2$.
This $2$ is related to the fact that the anomalous dependence on $\Omega$
we referred to earlier is ${\Omega}^{2-2g}$ on the genus $g$\footnote{The fields of the sigma model part of the {\bB} string are:
$x^{i}, {\bar x}^{\ib}, {\eta}^{\ib}, {\theta}_i , {\psi}^{i}_{\a}$,
so the unbalanced are $2g$ zero modes of the 1-form ${\psi}^{i}$, one zero
mode of $x^{i}$ and one of ${\theta}_i$. Since $\theta$ and $\psi$ are
fermions, their measure transforms as ${\Omega}^{1-2g}$, while the one of
the bosons $x^i$ gives another factor of $\Omega$}

The full topological string partition function includes also the higher
genus amplitudes:
\begin{equation} \label{topb}
Z_{B}(X^{\vee}; t, {\hbar}) = {\exp} \sum_{g=0}^{\infty}
{\hbar}^{2-2g} {\CF}_{g}(X^{\vee}; t)
\end{equation}
For small $\hbar$ the partition function behaves as:
$e^{{\CF}_{0}/{\hbar}^2 + \ldots}$
which is a quasiclassical expression for
a wave function, since
${\CF}_0$, thanks to (\ref{prep}), is a generating function of a Lagrangian
submanifold in $V = H^{3}(X^{\vee}, {\bR})$. The wave function, then,
corresponds to a state in the Hilbert space obtained by quantizing $V$.
However the Planck constant in this ``quantization'' is ${\hbar}^2$, not $\hbar$.
Moreover, \cite{bcov} has shown, that $Z_{B}$ cannot be viewed as a holomorphic
function of $t$. Instead, the naive decoupling of ${\bar t}$ dependence
is replaced by a certain linear partial differential equation on
$Z_B$, called the holomorphic anomaly equation \cite{bcov}, which was interpreted in \cite{wittenqbi} as an equation, expressing
the dependence of the wave function, obtained by the quantization of $V$, on
the choice of holomorphic polarization\footnote{There is some confusion, however, as to whether
it is $Z_B$ which has such an intepretation, or its square}.

Note that one has a lot of freedom in parameterizing $\widetilde\CM$. The $A$-periods
of $\Omega$ provide local holomorphic coordinates, but they may be not the
most
useful ones. The definition of these
coordinates required a choice of the basis in $H_{3}(X^{\vee})$ but,
since $H_{3}(X^{\vee},{\bR}) = H_{3}(X^{\vee}, {\bZ})\otimes {\bR}$
this choice, made for some particular
$X^{\vee}$, can be uniquely extended to all nearby CY's, and also globally up to monodromy in
 $Sp( 2r+2
,{\bZ})$.
In the holomorphic coordinates $t$ the prepotential
${\CF}_0$ makes its most natural appearence, but it could be that it is not
the most natural object to look at, especially in view of the holomorphic
anomaly. Take, for example, the real part of
$\Omega$, ${\Phi} = {\rm Re}{\Omega}$, and parameterize
$\widetilde\CM$ by the cohomology class $\varphi$ of $\Phi$.
In other words, let us pass from $t_i$ parameterization to
$(p_i , q^i )$ parameterization:
$p_i = \oint_{A^i} {\Phi} ,  q^{i} = \oint_{B_{i}} {\Phi}$.
Clearly, from (\ref{perds}) we can express $(p,q)$ via $t, {\bar t}$:
\begin{equation} \label{reltn}
p_i = {\half} \left( t_i + {\bar t}_i \right), \qquad
q^i = {\half} \left( {\p{\CF}_0 \over {\p}t_{i}} + {\p{\bar \CF}_0  \over
{\p}{\bar t}_i} \right)
\end{equation}
We claim that the transformation $(t, {\bar t}) \mapsto (p,q)$ is generated
by the generating function, which turns out to be quite natural from the
point of view of six dimensional topological gravity. In order to see that,
introduce one more notation:
$t_i = p_i + \sqrt{-1} {\xi}_i ,  i = 0, 1, \ldots, r$ and
consider the following function on $\widetilde\CM$:
\begin{equation} \label{vlm}
{\CV} = {1\over 2\sqrt{-1}}\int_{X^{\vee}} {\Omega} \wedge {\bar\Omega}
\end{equation}
In the effective
four dimensional supegravity obtained by compactifying Type {\bf II}
string on CY $X^{\vee}$ ${\CV}$ gives the exponential
of the Kahler potential.
We note in passing that the $(p,q)$ coordinates on ${\widetilde\CM}$ are
analogous to the Penner coordinates ${\ell}_i$ on the combinatorial
moduli space of Riemann surfaces, which are not holomorphic,  but are quite
useful\cite{maxim}. Thus their six
dimensional analogues are also natural to consider.
We can easily relate $\CV$ to ${\CF}_0$: $$2\sqrt{-1}{\CV} =  \sum_j \left( \oint_{A^j} {\Omega}
\oint_{B_j} {\bar \Omega} - \oint_{A^j} {\bar\Omega}
\oint_{B_j} {\Omega} \right) =
\sum_j \left( t_j {\p{\bar\CF}_0
\over {\p}{\bar t}^j} - {\bar t}_j {\p{\CF}_0
\over {\p} t^j} \right) = $$
$$
2\sqrt{-1}(  2 \sum_j q^j {\xi}_j - H), $$
\begin{equation} \quad
 H = {1\over \sqrt{-1}} \left( {\CF}_0 - {\bar \CF}_0 \right)
 \qquad\qquad 2 q^j = {{\p}H \over {\p}{\xi}_{j}}\label{vtoprep}
 \end{equation}
Thus, $\CV$ is the Legendre transform of $H$ with respect to
$\xi$ and it is more natural to view $\CV$ as a function of $p$ and $q$,
not as a function of $p$ and $\xi$. Note that the similar Legendre transform
(and its quantum analogue) arise in the black hole entropy counting of
\cite{osv}.

\noindent
{\bf Hitchin's approach.}
There exists an interesting variational problem, which produces
$\CV$ directly as a function of $p$ and $q$ \cite{hitchini}. We present a slightly
reformulated version of Hitchin's construction (it was found independently in
\cite{saman}). In what follows we shall call this theory {\bH6}, since we shall also have to
look at the higher dimensional versions of that theory.
The fields of {\bH6} theory are the closed three-form ${\Phi}$ on
$X^{\vee}$ and a traceless vector-valued one-form $J$, i.e. a section of $End(TX^{\vee},
TX^{\vee})$, in other words, a Higgs field acting on the (real!) tangent bundle to
$X^{\vee}$. The one-form $J$ is further constrained, which is implemented by yet
another field, a six-form ${\ve}$, which enters the functional linearly.
Here is the functional:
\begin{equation} \label{htich}
S_{\bH6} = \int_{X^{\vee}} {\Phi} \wedge \iota_{J} {\Phi} + i {\ve}
\left( {\Tr}J^2 + 6 \right) = \end{equation}
$\int_{X^{\vee}} {\Phi}_{[abc} J_{d}^{m} {\Phi}_{ef]m} ( {\rm
d}^6 x)^{abcdef} $,
provided that the constraints
\begin{equation} \label{cnstr}
J_{a}^{b}J_{b}^{a} = - 6, \quad J^{a}_{a}= 0
\end{equation}
are imposed. Finally, the real variable is not $\Phi$, but a two-form $B$,
which enters as follows:
fix ${\vf} = [ {\Phi} ] \in H^{3}(X^{\vee}, {\bR})$. We shall denote by the
same letter the de Rham representative of ${\vf}$, i.e. a closed $3$-form.
Then we write:
\begin{equation} \label{bfild}
{\Phi} = {\vf} + {\rm d}B, \qquad B \in {\Lambda}^{2}T^{*}X^{\vee}
\end{equation}
The claim of Hitchin's is that by minimizing $S$ with respect to $B$ one
gets $\CV$, where ${\Omega} = {\Phi}_{*} + {\sqrt{-1}}
\iota_{J_{*}}{\Phi}_{*}$, $J_{*}^2 = - 1$, and
$*$ means that we take the values of $\Phi$ and $J$ at the critical point of
$S$.
We note in passing that the Lagrangian, analogous to (\ref{htich}), can be
written in two dimensions, where one replaces $\Phi$ by a closed $1$-form.
In this case one gets Polyakov's formulation of the sigma model coupled to
two dimensional gravity, where, however, the two dimensional metric appears
only
through the complex structure $J$: $J_{\a}^{\b} =
\sqrt{g}g^{\b\g}{\e}_{\a\g}$. Minimizing with respect to $J$ is not very
sensible unless one considers not one, but at least two closed one-forms
${\Phi}^i$ (target space should be at least two dimensional, for the
classical Polyakov string to be equivalent to the Nambu string). Such a
generalization is also possible in six dimensional context \cite{saman}. Note, however, that
unlike two dimensional case,
one can only consider ``flat tensorial target spaces with constant
$\mathfrak B$-field'':
$S_{\mathfrak G, B} = \int {\mathfrak G}_{\a\b} {\Phi}^{\a} \wedge \iota_{J} {\Phi}^{\b}
+ {\mathfrak B}_{\a \b} {\Phi}^{\a} \wedge {\Phi}^{\b}$
where the index $\a$ runs from $1$ to $d$, the dimension of the
tensorial target space, and $\mathfrak G$ and $\mathfrak B$ are the constant metric and the
antisymmetric tensor respectively\footnote{For $d=2$ such theories
are
closely related to the theories on $(p,q)$ fivebranes in {\bf IIB} string
theory}.

In a sense, {\bH6}-theory exhibits T-duality,
just like two dimensional sigma model on a circle\footnote{The implementation of the T-duality is quite interesting in view of
applications to the black hole entropy counting \cite{osv}. One relaxes
the constraint that ${\Phi}$ is closed, and adds a term
$\int {\Phi} \wedge {\tilde\Phi}$ to the action, where ${\tilde\Phi}$ is
the closed three-form, with fixed cohomology class. Field theoretically
it is more natural (and, as we explain below, necessary after coupling to
the gauge fields) to assume that $[{\tilde\Phi}]\in H^{3}(X^{\vee}, {\bZ})$}. However, it maps
quadratic constraint ${\Tr}J^2 = -6$ into the nonlinear one: ${\Tr}J^{-2} =
-6$. At any rate, {\bH6} provides an interesting off-shell
extension of the prepotential.

\noindent
{\bf Kodaira-Spencer theory.}We don't know whether this is a
``correct'' off-shell extension, i.e. if it reproduces the higher genus amplitudes
${\CF}_{g}$, $g \geq 1$. It has been argued in \cite{bcov} that these ampltidues
can be calculated using the Feynmann rules of the so-called Kodaira-Spencer
(KS)
theory of gravity, which is a cubic field theory, whose propagating field is
a $(1,1)$ two-form on $X^{\vee}$. In this sense KS theory is similar to
{\bH6} theory, although in the latter the propagating field $B$ is just a
two-form. Both theories are background dependent: in KS case one has to fix the reference complex structure and an element in $H^{2,1}(X^{\vee})$;
in {\bH6} theory one has to fix $\varphi$.
We refer to \cite{saman} for the suggestion for the
construction of the map between the KS and
{\bH6} theories, although neither \cite{saman} nor we suggest that any of
these formulations are the suitable nonperturbative formulations of the
type {\bB} topological string.
We stress that {\bH6} theory, if at all, is
related to the ``square'' of the topological {\bB} string, since it is
${\CF}_0 - {\bar\CF}_0$ which is the Legendre transform of $H$, not the
prepotential itself. The same Legendre transform
occured in the recent discussion of black hole entropy \cite{osv}.

\noindent
{\bf Nonperturbative corrections.} Neither KS nor Hitchin's
functionals know about the non-perturbative corrections coming
from D-branes in the type {\bB} string. The latter correspond to the coherent
sheaves on $X^{\vee}$, and the simplest ones are the ideal sheaves of points and holomorphic curves on
$X^{\vee}$. The actual counting problem which their consideration leads to
will be described in the Donaldson-Thomas section. Here we shall simply
mention that their enumeration brings extra parameters the partition
function of the type {\bB} string must depend on: the two-form $s \in
H^{2}(X^{\vee}, {\bC})$, which couples to the worldvolumes of the D-strings,
the Poincare duals of ${\rm ch}_2$ of the corresponding sheaves:
\begin{equation} \label{nonperb}
Z_{B}^{?}(X^{\vee}; t, s, {\hbar} \vert {\Lambda}) =
Z_{B}(X^{\vee}; t) \times \sum_{{\l} \in  \Lambda \subset H^{even}
(X^{\vee}, {\bZ})}
{\exp} \left( {1\over{\hbar}} \int_{X^{\vee}} {\l} \wedge s \right)
\int_{[{\bf M}_{\l}]^{vir}} 1 \qquad
\end{equation}
where ${\bf M}_{\l}$ is the moduli space of (stable?) coherent sheaves
${\CI}$ on $X^{\vee}$, with the Chern character ${\rm ch}({\CI}) = {\l}$,
and ${\Lambda}$ is some Lagrangian sublattice\footnote{The most
studied, so far, example, corresponds to
$$
{\Lambda} = ( 1 \in H^{0}(X^{\vee}, {\bZ})) \oplus 0 \oplus  H^{4}(X^{\vee}, {\bZ})
\oplus  H^{6}(X^{\vee}, {\bZ})
$$} in $H^{even}$, and $[{\bf M}_{\l}]$ is the virtual fundamental cycle.

\noindent{\bf Donaldson-Thomas corner.}
The Donaldson-Thomas (DT) theory is the mathematical version of the physical
``integrating out the D-branes'' procedure. The theory is not yet constructed,
but some partial results are already available, especially in the rank one
case.
We consider Calabi-Yau threefold $Y$\footnote{We
choose here another letter for the CY space, since depending on the context
$Y$ may stand for $X$ or for $X^{\vee}$ in the discussions above}.
Just as topological string of type {\bA} allows generalizations to non-CY
spaces, the DT theory also has a non-CY version, however we shall not
discuss it here.
Type {\bB} open strings couple to
$(0,1)$-connections ${\bar A}$, which correspond to the $Q$-closed boundary operators
iff $F^{0,2}= 0$ \cite{wittencs}, the naive guess would be that we should look for
the solutions of the equations $F^{0,2} = 0$ modulo (complexified) gauge
transformations. This is the same thing as solving holomorphic Chern-Simons
(hCS)
equations of motion.
However, deformation-theoretically this is not a very well-posed problem.
Indeed, $\bar A$ has three functional degrees of freedom, $F^{0,2}= 0$
imposes three equations, and we have one gauge invariance. Therefore, the
virtual dimension of the space of solutions is minus infinity.
This problem can be, however, cured, by introducing the adjoint Higgs field
${\varpi}$ which is a $(0,3)$-form with values in the endomorphisms of the
bundle where $\bar A$  acts. The equations $F^{0,2}=0$ are replaced by the
so-called Donaldson-Uhlenbeck-Yau equations:
\begin{equation} \label{duy}
 F^{0,2}_{\bar A} = {\pb}^{\dagger}_{\bar A}{\varpi} , \qquad
 F^{1,1} \wedge k\wedge k = [ {\varpi}, {\bar\varpi}]
\end{equation}
where we have also partly fixed the gauge, leaving only the unitary gauge
transformations. This partial gauge fixing is the physical implementation of the stability
condition. It explicitly depends on the Kahler form $k$.
The equations (\ref{duy}) no longer follow from the hCS action. Instead, they
describe the localization locus of the (partially) topologically twisted ${\CN}=2$ six
dimensional gauge theory, which lives on the Euclidean D5-brane wrapping a
six-fold inside a CY fourfold. If sixfold
itself is a CY threefold, then ${\varpi}$ is a scalar, and
moreover on the solutions of (\ref{duy}) $\varpi$ vanishes,
formally reducing us to the original hCS problem. However, the presence of
$\varpi$ is important in evaluating the determinants of the fluctuations
around the solutions to (\ref{duy}) and, ultimately, in construction of the
virtual fundamental cycle of ${\bf M}_{\l}$.

The hCS theory has the Lagrangian, derived in the {\bB} string context in
\cite{wittencs}:
\begin{equation} \label{hcsth}
\int_{X^{\vee}} {\Omega} \wedge {\Tr} ( A dA +{2\over 3} A^{3})
\end{equation}
If we follow the previous philosophy and couple the {\bB} model
to ${\bar \bB}$ model, then we should replace (\ref{hcsth}) by:
\begin{equation} \label{hcsthh}
\int_{X^{\vee}} {\Phi} \wedge {\Tr} ( A dA +{2\over 3} A^{3})
= \int_{X^{\vee}} {\varphi} \wedge CS(A) +  B \wedge {\Tr} ( F \wedge F
)
\end{equation}
The theory with the action (\ref{hcsth}) makes almost no sense, since
the exponential of the action (\ref{hcsth}) is not gauge invariant. Indeed,
its gauge invariance requires integrality of $[ {\Omega} ] \in H^{3}(X, {\bC})$
but it is impossible to achieve for compact non-singular CY. The action
(\ref{hcsthh}), on the other hand, is perfectly sensible, since the requirement
of integrality of ${\varphi} = [ {\Phi} ]$ is quite a reasonable one.
Moreover, this requirement is {\it precisely} the condition \cite{mooreattr} on the
complex moduli of CY to be the solution of {\it attractor equations}.
We see therefore that the topological strings know something about the
black holes, constructed by wrapping D-branes on various cycles in the CY.
Perhaps this remark would clarify some of the mysteries uncovered in \cite{osv}.

The last term in (\ref{hcsthh}) couples $B$ to the second Chern class of the bundle. In
general we should consider not just bundles and connections, but the
complexes of bundles, with connections and maps (bi-fundamental matter)
between  them. The notion of connection on the object in derived category
gets complicated but the Chern character and its component are still simple.

In fact, even if the complex of bundles corresponds to the ideal sheaf
of a curve or of a collection of curves and points, which are the objects
of study in the DT theory, the second Chern class has a simple meaning.
It is the Poincare dual to the cycles, represented by the curves. The
coupling
(\ref{hcsthh}) describes, then, the coupling of the $B$-field to these curves.
We thus learn that the $B$-field of {\bH6} theory plays the role of the Kahler
form for the D-strings of the corresponding type {\bB} topological string!

\noindent
{\bf DT theory and quantum space-time foam.}
The discovery of \cite{ionv} was the realization that the counting of
ideal sheaves \cite{mnop}, which is performed by the DT theory can be viewed as
the Kahler gravity path integral, where one sums over fluctuating topologies
of the six dimensional space-time. In this interpretation, the ``curvature''
$F$ which is used to represent the Chern classes of the sheaves, is viewed
as the (discrete) deformation of the Kahler form: $k = k_{0} + {\hbar}F$.
We should stress that the results of \cite{ionv} do not imply the
discreteness of the fundamental description of Kahler gravity. We do not
know what is the correct Lagrangian of that theory, nor what are its
fundamental degrees of freedom. However, the localization technique can be
applied to this theory, and the fixed points of the symmetry group action
(the symmetry in question was the torus action, which was the isometry
of the background toric CY, and acted on the space of Kahler metrics which
asymptote the background one) corresponded to the blowups of the original CY
along the ideals of the torus invariant curves.
The visible part of the Kahler gravity action is the volume of the space:
\begin{equation} \label{kahla}
S_{\rm kahler} = {1\over 6 {\hbar}^2}\int k \wedge k \wedge k =
S_{0} + \int k_{0} \wedge {\rm ch}_2 + {\hbar} \int {\rm ch}_{3}
\end{equation}
As explained in \cite{nov} the equality of the DT and GW partition functions
is the particular case of more general phenomenon -- S-duality which relates
{\bA} and {\bB} topological strings on the same CY manifold, while inverting
the string coupling $\hbar$. In the case of toric CY's the KS contribution
to the {\bB}-partition function is trivial, as the D-brane contribution to
the
{\bA}-partition function. The S-duality can be derived from the S-duality
of physical {\bf IIB} string \cite{nov}, \cite{kapustin}.

\subsection{Donaldson-Witten corner}

$\bZ$-theory also knows about four dimensional gauge theory. Fix a gauge
group $G$. Then one can define a partition function:
\begin{equation}
Z ( {\mathfrak a}, {\ve}_{1} , {\ve}_{2} , q ) =  Z^{pert}( {\mathfrak a}, {\ve}_1 , {\ve}_2 , q )
\times \sum_{n=0}^{\infty} q^{2 h^{\vee}_{G} n}\
{\Tr}_{{\bC}[{\bf M}_{n}(G)]} \left( {\exp}({\mathfrak a}), e^{{\ve}_{1}},
e^{{\ve}_{2}} \right)\qquad\label{omg}
\end{equation}
where
${\bf M}_{n}(G)$ denotes the moduli space of framed holomorphic
$G_{\bC}$-bundles ${\CP}$ over ${\bC\bP}^2 = {\bC}^{2} \cup {\bC\bP}^1_{\infty}$, which are trivial on
${\bC\bP}^1_{\infty}$, together with the choice of the trivialization:
$$
{\CP}_{{\bC\bP}^1_{\infty}} \approx {\bC\bP}^1_{\infty} \times G_{\bC}
$$
The number $n$ is the second Chern class $c_2({\CP})$ of the bundles,
${\exp}{\mathfrak a} \in {\bT}$ -- the maximal torus of $G_{\bC}$, $(e^{{\ve}_1},
e^{{\ve}_2}) \in {\bC}^{*} \times {\bC}^{*}$ -- the maximal torus of
$SO_{4}({\bC})$ -- the complexification of the of group of isometries of
${\bR}^{4} \approx {\bC}^2$, and finally ${\bC}[X]$ denotes the space of
holomorphic functions on $X$ (for the classical gauge groups from the
A,B,C,D seria one can think of the polynomials in the ADHM data).
The group $G_{\bC} \times SO_{4}$ acts on ${\bf M}_{n}(G)$ by changing
the trivialization at infinity and by
automorphisms of ${\bC\bP}^2$, preserving infinity. Finally, the $Z^{pert}$
is the perturbative (from the gauge theory point of view) piece, which
can be found in \cite{abcd}.
The trace in
(\ref{omg}) arises in the counting of the BPS states in the five dimensional gauge
theory compactified on a circle \cite{fivedim}. This gauge theory
can be engineered \cite{vafaengine} by
compactifying {\bM}-theory on local Calabi-Yau manifold $X_{G}$, which is the
$A,D,E$ singularity fibered over ${\bP}^1$ (for non-simply laced $G$ one
also twists by the automorphisms of the Dynkin diagrams of the corresponding
$A,D,E$ ``parent'' groups).
The correspondence between physical {\bM}-theory compactification and the
low-energy physics of the resulting four dimensional gauge theory implies
that the same $Z$ function should be equal to the Gromov-Witten partition
function of $X_{G}$, where $\mathfrak a$ and ${\rm log}(q)$ correspond to the
Kahler moduli of $X_{G}$ \cite{nekinst}. The parameters ${\ve}_1, {\ve}_2$ are trickier
to identify. In the simplest specialization, when ${\ve}_1 = - {\ve}_2 =
{\hbar}$, the latter is identified with the string coupling constant.
However, general case is harder to intepret. Presumably it corresponds
to the equivariant GW (= equivariant DT) theory of $X_{G}$.

\section{Towards {\bZ}-theory}

\hfill{\vbox{\hbox{\tiny - Who's Zed?}\hbox{\tiny - Zed's dead, baby, Zed's dead.}}}

In our discussions of GW, KS, DT corners of $\bZ$-theory we arrived at the
picture where the complete(d) topological string partition function
depends on both  $\mathfrak s$ and $\mathfrak t$ variables, or $t$ and $s$ variables.
It is plain to identify them with the {\it full moduli} of CY
metric on $Y$. Nonperturbative topological string cares
 about both the calibrated complex and symplectic aspects of CY
geometry. We are not saying
that the exact CY metric is what the string couples to. Rather,
it is the {\it homological CY geometry} that the topological string
cares about. One way to make the unification between the Kahler
and calibrated complex moduli of $Y$ is to consider the manifold
${\CZ} = Y \times
{\bS}^1$. Its third cohomology splits as $H^{2}(Y) \oplus H^{3}(Y)$.
In this way our moduli are nothing but the moduli of three-forms in
seven dimensional theory. Moreover, the Lagrangian branes and topological
{\bA} strings are nothing but the associative cycles in the $G_{2}$-manifold
${\CZ}$\footnote{In
his lecture at Les Houches School in the summer of 2001
S.~Shatashvili proposed the unification of the topological {\bA} and {\bB}
models as the main motivation for
the study of the analogue of the topological twist in the
Type {\bf II} superstring compactifications on $G_{2}$ and $Spin(7)$
holonomy manifolds\cite{samsonvafa}}.
\subsection{Hitchin theory in seven dimensions: Polyakov formulation}

\hfill{\vbox{\hbox{\tiny Agent 007 reports to M.}
}}

Hitchin has proposed \cite{hitchini} a seven dimensional theory ({\bH7}),
analogous
to {\bH6}, which classifies $G_2$-holonomy metrics on ${\CZ}$
in terms of the closed three-forms on ${\CZ}$. We present here its
Polyakov-like formulation. The fields of the {\bH7} theory are: the metric
$h_{ij}$ and a closed three-form $C = C_{ijk}{\dd y}^i {\dd y}^j {\dd y}^k$. The Lagrangian is:
\begin{equation}
S_{\bH7} = \int_{\CZ} \left( h^{ij} C_{i} \wedge C_{j} \wedge C + \sqrt{h}
\right)
\label{htichseven}
\end{equation}
where $C_{i} = C_{ijk} {\dd}y^j \wedge {\dd}y^k$. Again, the dynamical field
is not $C$ but the two-form $B$, such that $C = {\s} + {\dd}B$, and ${\s} = [ C ] \in
H^{3}({\CZ}, {\bR})$ is fixed.
Minimizing (\ref{htichseven}) with respect to $h$ one arrives at the
Lagrangian, proposed by N.~Hitchin in \cite{hitchini}. Note that there exists a T-dual version
of (\ref{htichseven}) which is non-polynomial in the propagating three-form.
As in the six dimensional case, one can treat $C$ as an independent field,
and add a term $\int C \wedge {\tilde G}$ to the action, where ${\tilde G}$
is a closed 4-form, with fixed cohomology class $[ {\tilde G} ]$. Then
classically one can eliminate $C$ and arrive at the action involving
${\tilde G}$ and $h$. Further eliminating $h$, one arrives at the dual
formulation of the theory, also present in \cite{hitchini}.

What is the meaning of the expression (\ref{htichseven})? We found the
following simple way of repackaging it.
Introduce Dirac matrices ${\g}^i$, which obey
$$
{\g}^{i} {\g}^{j} + {\g}^{j} {\g}^{i} = 2 h^{ij} \cdot {\bf 1}
$$
Then (\ref{htichseven}) can be rewritten as:
\begin{equation}
S_{\bH7} = \int_{\CZ} \sqrt{h} \left( 1 + {2\over 3} {\Tr} {\hat C}_{(3)}{\hat C}_{(3)}{\hat C}_{(3)} \right)
\label{spinhs}
\end{equation}
where ${\hat C}_{(3)} = C_{ijk} {\g}^{[i}{\g}^{j}{\g}^{k]}$ and
the trace is taken in the spin representation of $Spin(7)$.
We note that on the solutions of (\ref{htichseven}) the three-form $C$ is
harmonic with respect to the metric $h$ (which in turn depends on ${\s}$).
This condition can be neatly expressed as:
\begin{equation}
\{ {\dir}, {\hat C}_{(3)} \} = 0
\label{dirac}
\end{equation}
where ${\dir}$ is the Dirac operator.

The theory (\ref{htichseven}) on ${\CZ} =
{\bS}^1 \times Y$ on ${\bS}^1$-invariant fields reduces to the sum of the
Kahler gravity and the {\bH6} theory (although the constraint ${\Tr}J^2 =
-6$ is replaced by ${\Det}J = -1$, and one has to integrate out the dilaton
and the KK vector field). This makes the theory (\ref{htichseven}) a
suggestive candidate for the correct theory. However, this on-shell verification
is not sufficient for our purposes.

\subsection{Speculations on unification}

The action (\ref{spinhs})
involves the metric $h_{ij}$ and integrating it out seems as
difficult, as it is in four dimensions. We better use the lesson of
\cite{ionv} and replace the physical metric by the gauge fields.
We know in the case  ${\CZ} = {\bS}^1 \times Y$  that the
quantum foam picture is defined (and completed) quantum mechanically by
summing over holomorhic curves, which are the worldsheets of D-strings with
D(-1)-instantons bound to them. Lifted to seven dimensions this summation
becomes the summation over associative cycles. These are cycles for which
the
volume form, obtained from the induced metric, coincides with the
restriction of the three form $C$, which should solve the equations of
motion of (\ref{htichseven}).

We also should not forget about the Chern-Simons theory $ \int C dC$ of 3-forms in seven
dimensions. After all, it is this theory which explains in a most natural
way the holomorphic anomaly equation \cite{wittenqbi}, \cite{saman}.
However this theory does not include the D-brane effects in the topological
string. So we must supplement it with the contribution of membranes, which
are charged under the $C$-field:
\begin{equation}
S_{\rm ?} = {\half} \int_{\CZ} C {\dd} C + \sum_{{\l} \in H_{3}({\CZ}, {\bZ})}
{\CN}_{\l} e^{-\int_{\l} C}
\label{csmem}
\end{equation}
where ${\CN}_{\l}$ is the number of associative cycles in the homology class
$\l$. Indeed, the associative cycles would project to the holomorphic curves
upon reduction to six dimensions. However, the notion of associativity
requires the $G_2$ structure, which is only avalable (if at all) for a
special gauge representative of $C$. Also, the expression (\ref{csmem})
cannot possibly qualify for the definition of fundamental theory -- it is at
best some sort of effective action. The analogy we have in mind is the three
dimensional compact electrodynamics of Polyakov \cite{polyakov}, where
one writes the effective $U(1)$ theory contains instanton corrections, which
reflect the presence of some more fundamental degrees of freedom (for
example, non-abelian gauge fields).

If we invoke the instanton interpretation of the membranes in (\ref{csmem}),
we could hope to reproduce the sum over $\l$ from some sort of gauge theory
on ${\CZ}$, with gauge field $A$ interacting with $C$ via the coupling:
\begin{equation}
{\CS}_{\rm ??} = {\half}\int C {\dd} C + \int C \wedge {\Tr} F \wedge F
+\int CS_{7}(A)
\label{csgauge}
\end{equation}
The couplings in (\ref{csgauge}) can be shown to reproduce the on-shell values of the
action of Kahler
gravity (\ref{kahla}). However, (\ref{csgauge}) does not contain terms which
would ensure localization of the gauge theory path integral. For special gauge groups,
like $U(1)$ or $SO(8)$ or $E_{8}$ the variations
of the gauge field $A \mapsto A + {\d}A$ can be compensated by the
Green-Schwarz-like variation of the three-form: $C \mapsto C + {\Tr}({\d}A
F)$, so the theory has an enormous gauge invariance
as far as $A$ is concerned\footnote{In fact, for these gauge groups
by redefining $C \mapsto C + CS_{3}(A)$ one can eliminate $A$ dependence in the trivial instanton sector}.
The gauge-fixed theory could be defined
along the lines of \cite{kanno},\cite{btdia}
using the three-form $C$, which does not have to solve (\ref{htichseven}),
thus suggesting an off-shell extension.
Perhaps it is worth observing that (\ref{csgauge}) can also be interpreted as
a Chern-Simons action for the superconnection ${\CA} = A + C$:
\begin{equation}
{\dd}^{-1} {\Tr} e^{{\CF}}
\label{cssuper}
\end{equation}
where ${\CF} = {\dd}{\CA} + {\CA}^2$.

As to the {\bH7} theory, we believe that the ${\Tr}{\hat C}_{(3)}^3$
representation should be taken as hint to what the correct formulation of
the theory should be. We suggest to study the spectral action \cite{connes},
associated
with the generalized Dirac operator:
\begin{equation}
{\hat D} = {\dir} + {\hat C}
\label{gendir}
\end{equation}
where ${\hat C} = \sum_{p=0}^{3} {\hat C}_{(2p+1)}, {\hat C}_{(2p+1)} = C_{i_{1}i_{2}\ldots i_{2p+1}} {\g}^{[i_{1}}{\g}^{i_{2}}\ldots
{\g}^{i_{2p+1}]}$. Consider the following generalization of the trace of
heat kernel
\begin{equation}
I (t) =
{\Tr}_{\CH} e^{- t {\hat D}^2}
\label{spectr}
\end{equation}
where ${\CH}$ is the space of sections of spin bundle over ${\CZ}$ (we neglect here the subtleties
about the existence of spin structure).
The small $t$ expansion of $I(t)$ gives the functionals of $C_{(p)}$ and the
metric $h$, which are invariant under the following gauge transformations:
\begin{equation}
\left( {\dir} + {\hat C} \right) \mapsto e^{-{\hat B}} \left(
 {\dir} + {\hat C} \right) e^{{\hat B}}
\label{ggeinv}
\end{equation}
where ${\hat B}  = \sum_{p=0}^{3} {\hat B}_{(2p)}, {\hat B}_{(2p)} = B_{i_{1}i_{2}\ldots i_{2p+1}} {\g}^{[i_{1}}{\g}^{i_{2}}\ldots
{\g}^{i_{2p}]}$, so that ultimately  one should take ${\hat C}$ to contain all forms,
while reversing the statistics of even ones, as they would correspond to the
ghosts.
One can also consider the eight-dimensional index density,
$$
\int_{0}^{1} dt \ {\rm Lim}_{{\b}\to 0} \
{\Tr}\ {\g}^{9}\  e^{-{\b}( {\g}^8 {\p}_t + {\dir} + t {\hat C} )^2} ;$$
and extract the corresponding Chern-Simons-like action:
\begin{equation}
{\CS}_{\rm ???} = \int_{\CZ} \sqrt{h} \left( {\Tr} \left( {\hat C} \{ {\dir},  {\hat C} \} + {2\over 3} {\hat
C}^3 \right) + 1 \right)
\label{csspin}
\end{equation}
Varying (\ref{csspin}) with respect to
${\hat C}_{(5)}$  we get equations
on the gauge field ${\hat C}_{(1)}$, similar to (\ref{duy}).
What is left is the sum of
the
 $C dC$ Chern-Simons action and {\bH7}-action (\ref{htichseven}).

It would be obviously interesting to derive the actions (\ref{csgauge}),(\ref{csspin}) by:

i)
integrating out free fields (like free fermions in 3d), which might be
related to the gravitino of physical {\bM}-theory (see \cite{moorefreed} for the recent discussions of the subtleties of the latter, also
see \cite{saman} for alternative suggestions for the free field formulation);

ii)
from some sort of
topological open membrane theory \cite{pioline}.

The action (\ref{csspin}) might be also related to the recent
studies of flux compactifications and generalized CY manifolds \cite{x}, \cite{hitchinii}.
Perhaps the gauge field is nothing
but the component ${\hat C}_{(1)}$, and we have to include higher
Chern-Simons terms in the action (\ref{csspin}). It is also possible that the components
${\hat C}_{(5)}, {\hat C}_{(7)}$ should be viewed as BV anti-fields, and
should be gauge-fixed.

We also mention another connection to the noncommutative geometry. Obviously, the expression
(\ref{htichseven}) reminds very much the star product of differential forms,
viewed as functions on $\Pi T{\CZ}$,
where the non(anti)commutativity is introduced along the odd directions. It is plausible that one can write some sort of matrix model, where
${\g}^i$'s will be independent variables, so that the Dirac anticommutation relations would correspond to the minima
of the action. In this way
the metric $h^{ij}$ will be just a parameter of the classical solution, and not a fundamental field.
The analogous phenomenon in the contenxt of noncommutativity is well-known \cite{ncopen}.

Obviously, all this
deserves further investigation\footnote{As we were submitting this paper to the archive, an interesting
paper \cite{baulieu} appeared, which might also prove useful in the construction of the {\bZ}-theory}.

As one of the indications of the naturalness of the seven dimensional theory we give here the
formula for the partition function of this theory in the $\Omega$-background
on ${\bS}^1 \times {\bR}^6$, which is a generalization of the equivariant
MacMahon function \cite{mnop}:
\begin{equation}
\label{instprtni}
Z_{\rm 7d } =  \prod_{a,b=1}^{\infty} {{(1 - q_{+}^{a}
(q_{1}q_2 q_3)^{b-1})}\over{(1-q_{-}^a (q_1 q_2 q_3)^{b-1})}}\prod_{\a = 1}^{3} \left[ {{(1 - q_{-}^{a}
q_{\a}^{b})}\over{(1-q_{+}^a q_{\a}^b)}} \right]
\end{equation}
where $q_{1}, q_{2}, q_{3}$ are the equivariant parameters, and $q_{\pm} =
 q
(q_{1}q_{2}q_{3})^{\pm \half}$, $q = - e^{i \hbar}$. The partition function is the (conjectural)
answer for the sum over 3d partitions with the K-theoretic analogue of the
equivariant vertex measure. The cohomological version of (\ref{instprtni}), which corresponds to
the six dimensional DT theory, is known to be true \cite{mnop}.

{\bf Zed and two notes}. The seven dimensional theory is not the final word. The
Chern-Simons action (\ref{csspin}) clearly suggests eight-dimensional
Donaldson-like theory, whose boundary action would be (\ref{csspin}). The
equations (\ref{duy}), due to the extra field $\varpi$, also are most
naturally interpreted in the eight dimensional terms. Moreover,
(\ref{instprtni}) exhibits most symmetries when
written in terms of four equivariant
parameters,  $q_{4} = (q_{1}q_{2}q_{3})^{-1}$:
\begin{equation}
\label{instprtnii}
Z_{\rm 7\to 8} =   \prod_{a,b=1}^{\infty} \prod_{{\m} = 1}^{4} \left[ {{(1 - q_{-}^{a}
q_{\m}^{b})}\over{(1-q_{+}^a q_{\m}^b)}} \right]
\end{equation}

\section*{Acknowledgements}

I thank the organizers of String'04 for giving me the honour of presenting
my thoughts there. Lectures at Barcelona RTN winter school, ICTP spring school, and
Northwestern University spring
conference on mathematics of string theory were quite useful for
shaping my ideas. I thank the organizers and the participants of all these meetings.
I am grateful to A.~Losev, A.~Okounkov and S.~Shatashvili
for numerous useful discussions. I would also like to thank J.H.~Park and O.~Ruchaysky
for their interest and discussions. It was K.~Saraikin who brought Hitchin's Lagrangians
to my attention during Les Houches summer school in 2001.
Research was partly supported by {\cyr RFFI} grant 03-02-17554
and
by the grant {\cyr NSh}-1999.2003.2 for scientific schools, also by
RTN under the contract 005104 ForcesUniverse.
Part of the research was done while I was visiting Trinity College, Dublin,
SUNY Stony Brook  (II Simons Workshop),
NHETC at Rutgers University, IAS at Princeton, and RIMS at Kyoto University.
I thank all these institutions for their hospitality.

\end{document}